\documentclass[aps,prb,twocolumn,showpacs]{revtex4}
\usepackage{graphicx}
\usepackage{amsmath}
\usepackage{amssymb}

\begin{document}

\draft

\title{Magnetic properties of charged spin-1 Bose gases with ferromagnetic coupling}

\author{Jihong Qin\footnote[1]{Corresponding author, E-mail: jhqin@sas.ustb.edu.cn}, Xiaoling Jian and Qiang Gu}

\address{Department of Physics, University of Science and Technology
Beijing, Beijing 100083, China}

\begin{abstract}
Magnetic properties of a charged spin-1 Bose gas with ferromagnetic
interactions is investigated within mean-field theory. It is shown
that a competition between paramagnetism, diamagnetism and
ferromagnetism exists in this system. It is shown that diamagnetism,
being concerned with spontaneous magnetization, cannot exceed
ferromagnetism in very weak magnetic field. The critical value of
reduced ferromagnetic coupling of paramagnetic phase to
ferromagnetic phase transition $\overline I_{c}$ increases with
increasing temperature. The Lande-factor $g$ is introduced to
describe the strength of paramagnetic effect which comes from the
spin degree of freedom. The magnetization density $\overline M$
increases monotonically with $g$ for fixed reduced ferromagnetic
coupling $\overline I$ as $\overline I>\overline I_{c}$. In a weak
magnetic field, ferromagnetism makes immense contribution to the
magnetization density. While at a high magnetic field, the
diamagnetism inclines to saturate. Evidence for condensation can be
seen in the magnetization density at weak magnetic field.
\end{abstract}


\maketitle

\section{Introduction}

The magnetism of Fermi gases has always received considerable
attention in solid-state physics, such as localized and itinerant
electrons. While the magnetic properties of Bose gases has been less
studied. But since the realization of Bose-Einstein condensation
(BEC) in ultracold atomic gases \cite{Anderson}, more interests have
been cast to this system. The Bose gases plays an important role in
understanding some exotic quantum phenomena, such as
superconductivity and superfluid. The ideal charged bosons were used
originally to describe the superconductivity. It has been shown by
Schafroth \cite{Schafroth}, Blatt and Butler \cite{Blatt} that an
ideal gas of charged bosons exhibits the essential equilibrium
features of superconductor. Although the Bardeen-Cooper-Schrieffer
(BCS) theory \cite{Bardeen} explained the microscopic nature of
conventional superconductivity, the charged Bose gas exhibits strong
diamagnetism at low temperature, which can be attributed to Meissner
effect. In recent years, the normal-state diamagnetism of
high-temperature cuprate superconductors has been explained by
real-space charged bosons \cite{Alexandrov}. This also recasts new
research interest in charged Bose gases.

Experimentally, since the realization of \emph{spinor} BEC in
optical traps \cite{Stamper-Kurn,Stenger} the magnetic properties of
\emph{spinor} Bose gases has received considerable attention.
Moreover, an ultracold plasma can be created by photoionization of
laser-cooled neutral atoms \cite{Killian}. The temperatures of
electrons and ions can reach as low as 100 mK and 10 $\mu$K,
respectively. The ions can be regarded as charged bosons if their
spins are integers. The Lande-factor for different magnetic ions
could also be different.

It is known that paramagnetism is from the spin degree of freedom of
particles. While charged spinless Bose gases can exhibit strong
diamagnetism, similar to Meissner effect, which comes from the
orbital motion of charge degree of freedom in magnetic field.
Theoretically, both the paramagnetism \cite{Yamada,Simkin} in
neutral spin-1 Bose gases and the diamagnetism of the charged
spinless Bose gases \cite{Daicic,Toms} have been studied. Moreover,
we \cite{Jian} have discussed the competition of paramagnetism and
diamagnetism in charged spin-1 Bose gases in external magnetic
field, using the Lande-factor $g$ to evaluate the strength of
paramagnetic (PM) effect. It is shown that the gas exhibits a shift
from diamagnetism to paramagnetism as $g$ increases.

The ferromagnetism and superconductivity are not compatible in
conventional physical models. The Meissner-Ochsenfeld effect shows
the conventional superconductor cancels all magnetic field inside
when the temperature below the superconducting transition
temperature, which means they become perfectly diamagnetic. The
discovery of several ferromagnetic (FM) superconductors in
experiments \cite{Saxena,Aoki,Slooten} stimulates the research
interest in the exotic magnetic properties of FM superconductors.
The state of the Cooper pairs in the FM superconductors has been
wildly studied \cite{Saxena,Aoki,Slooten,Machida,Monthoux}. A
stronger spin-orbit interaction in UGe$_{2}$ results in an abnormal
huge magnetocrystalline anisotropy \cite{Saxena,Aoki,Slooten}.
Monthoux et al.\cite{Monthoux} indicates that the favorite
superconducting pairing type of this anisotropy is triplet. Although
the exact symmetry of the paired state has not yet been identified,
a spin-triplet pairing is more likely than the spin-singlet pairing
in these superconductors \cite{Saxena,Aoki,Slooten}. These behaviors
are somewhat like charged spin-1 bosons. Thus the charged spin-1
boson model helps to understand the exotic magnetic properties
observed in such materials.

Although the ferromagnetism \cite{Ho,Ohmi,Gu,Tao,Kis-Szabo,Ashhab}
in a chargeless spinor Bose gas has also been involved in theory, it
is little discussed when FM interaction exists in a charged spin
system. Accordingly the magnetic behavior will become more complex
in charged spin systems with FM interactions, where diamagnetism,
paramagnetism and ferromagnetism compete with each other in such
case.

In this paper, the magnetic properties of a charged spin-1 Bose gas
with FM interactions are studied via mean-field theory.  Alexandrov
et al. found that the Coulomb or any other scattering may make
charged Bose gases superconducting below a critical field
\cite{Alexandrov1} with a specific vortex matter \cite{
Alexandrov2}. Superconducting is not obtained in our paper, probably
because we used the mean-field approximation to deal with the FM
interaction. In despite of this, mean-field theory is still
effective to point out the main physics of the magnetism, especially
the ferromagnetic transition \cite{Gu}. The remainder of this paper
is structured as follows. In Section 2, we construct a model
including Landau diamagnetism, Pauli paramagnetism and FM effect.
The magnetization density is obtained through the analytical
derivation. In Section 3, the results is obtained and the
discussions of our results is presented. A summary is given in
Section 4.

\section{The model}

The spin-1 Bose gas with FM couplings is described by the following
Hamiltonian:
\begin{eqnarray}\label{Hamil}
H - \mu {N}= D_L\sum_{j,k_z,\sigma}\left(\epsilon^l_{jk_{z}}
+\epsilon_{\sigma}^{ze}+\epsilon_{\sigma}^{m}-
\mu\right)n_{jk_z\sigma},
\end{eqnarray}
where $\mu$ is the chemical potential and the Landau levels of
bosons with charge $q$ and mass $m^{\ast}$ in the effective magnetic
field $B$ is
\begin{eqnarray}\label{Diam}
\epsilon^l_{jk_{z}}=(j+\frac{1}{2})\hbar \omega + \frac{\hbar^{2}
k_{z}^{2}}{2m^{\ast}},
\end{eqnarray}
where $j=0,1,2,\ldots$ labels different Landau levels and $ \omega
=qB/(m^{\ast}c)$ is the gyromagnetic frequency. The energy level is
degenerate with degeneracy
\begin{eqnarray}\label{Deg}
D_L=\frac{qBL_{x}L_{y}}{2\pi\hbar c},
\end{eqnarray}
where $L_{x}$ and $L_{y}$ are the length in $x$ and $y$ directions
of the system, respectively. The intrinsic magnetic moment
associated with the spin degree of freedom leads to the Zeeman
energy levels split in the magnetic field,
\begin{eqnarray}\label{Param}
\epsilon_{\sigma}^{ze} =  -  g\frac {\hbar q}{m^*c} \sigma B,
\end{eqnarray}
where $g$ is the Lande-factor and $\sigma$ denotes the spin-z index
of Zeeman state $\left| {F=1,m_F=\sigma} \right\rangle$ ($\sigma= 1,
0, -1$). The contribution to the effective Hamiltonian from the FM
couplings is
\begin{eqnarray}\label{Ferrom}
\epsilon_{\sigma}^{m} =  - 2 I\sigma( m+\sigma n_{\sigma}),
\end{eqnarray}
where $ I$ denotes FM coupling and spin polarization $ m= n_{1}-
n_{-1}$. The grand thermodynamic potential is expressed as
\begin{eqnarray}\label{T1}
\Omega_{T\neq0}& =&-\frac{1}{\beta}\ln\mathrm{Tr}e^{-\beta(H- \mu N)}\nonumber \\
&=&\frac{1}{\beta}D_L\sum_{j,k_z,\sigma}\ln
[1-e^{-\beta(\epsilon^l_{jk_{z}}
+\epsilon_{\sigma}^{ze}+\epsilon_{\sigma}^{m}- \mu )}],
\end{eqnarray}
where $\beta=(k_{B}T)^{-1}$. Through converting the sum over $k_z$
to continuum integral, we obtain
\begin{eqnarray}\label{T3}
\Omega_{T\neq0}&=&\frac{ \omega
m^{\ast}V}{(2\pi)^{2}\hbar\beta}\sum_{j=0}^{\infty}\sum_{\sigma}\int dk_z\nonumber\\
&\times&\ln\{1-e^{-\beta[(j+\frac{1}{2})\hbar
\omega+\frac{\hbar^{2}k_z^{2}}{2m^{\ast}}
 -g\frac {\hbar q}{m^*c} \sigma B- 2 I\sigma( m+\sigma  n_{\sigma})- \mu]}\},\nonumber\\
\end{eqnarray}
where $V$ is the volume of the system. Eq. (\ref{T3}) can be
evaluated by Taylor expansion, and then performing the integral over
$k_{z}$. We get
\begin{eqnarray}\label{T2}
\Omega_{T\neq0}&=&-\frac{ \omega
V}{\hbar^{2}}\left(\frac{m^{\ast}}{2\pi\beta}\right)^{3/2}\nonumber\\
&\times&\sum_{l=1}^{\infty}\sum_{\sigma}\frac{l^{-\frac{3}{2}}e^{-l\beta[\frac{\hbar
\omega}{2}
 -g\frac{\hbar q}{m^{\ast}c} \sigma
B- 2 I\sigma( m+\sigma  n_{\sigma})- \mu]}}{1-e^{-l\beta\hbar
\omega}}.\nonumber\\
\end{eqnarray}
For convenience's sake, we introduce some compact notation for the
class of sums. It can be defined as
\begin{eqnarray}\label{s}
\Sigma_{\kappa\sigma}[\alpha,\delta]=\sum_{l=1}^{\infty}\frac{l^{\alpha/2}e^{-l
x(\varepsilon+\delta)}}{(1-e^{-l x})^{\kappa}},
\end{eqnarray}
where $ x=\beta\hbar \omega$ and $
\mu-\epsilon_{\sigma}^{ze}-\epsilon_{\sigma}^{m}
=(\frac{1}{2}-\varepsilon)\hbar \omega$. Within this notation, Eq.
(\ref{T2}) can be rewritten as
\begin{eqnarray}\label{Ts}
\Omega_{T\neq0}=-\frac{ \omega
V}{\hbar^{2}}\left(\frac{m^{\ast}}{2\pi\beta}\right)^{3/2}\sum_{\sigma}\Sigma_{1\sigma}[-D,0].
\end{eqnarray}
with $D=3$. The particle density $ n=N/V$ can be expressed as
\begin{eqnarray}\label{n}
 n_{T\neq 0}&=-\frac{1}{V}\left(\frac{\partial\Omega_{T\neq 0}}{\partial\mu}\right)_{T,V}\nonumber\\
&=
x\left(\frac{m^{\ast}}{2\pi\beta\hbar^{2}}\right)^{3/2}\sum_{\sigma}\Sigma_{1\sigma}[2-D,0].
\end{eqnarray}
The magnetization density $ M$ can be obtained from the grand
thermodynamic potential,
\begin{eqnarray}\label{Mag}
 M_{T\neq 0} &=&-\frac{1}{V}\left(\frac{\partial\Omega_{T\neq 0}}{\partial B}\right)_{T,V}\nonumber\\
&=&\frac{\hbar q}{m^{\ast}c}\left(\frac{m^{\ast}}{2\pi\beta\hbar^{2}}\right)^{3/2}\sum_{\sigma}\biggl\{\Sigma_{1\sigma}[-D,0] \nonumber\\
&+& x(g\sigma-\frac{1}{2}) \Sigma_{1\sigma}[2-D,0]-
x\Sigma_{2\sigma}[2-D,1]\biggr\}.\nonumber\\
\end{eqnarray}
The relation among effective magnetic field $B$, external magnetic
field $H$ and magnetization density $M$ is formally expressed as
\begin{eqnarray}\label{mag}
B=H+4\pi M,
\end{eqnarray}
For computational convenience, some dimensionless parameters are
introduced below. $t=T/T^{\ast}$, $\overline M=m^{\ast}c{M}/( n\hbar
q)$, $\overline \omega=\hbar {\omega}/(k_{B}T^{\ast})$, $\overline
I= I n/(k_{B}T^{\ast})$,$\overline \mu= \mu/(k_{B}T^{\ast})$,
$\overline m= m/ n$, $\overline n_{\sigma}= n_{\sigma}/ n$ and
$h=\hbar qH/(m^{\ast}c k_{B}T^{\ast})$, and then $x=\overline
\omega/t$, where $T^{\ast}$ is the characteristic temperature of the
system, which is given by $k_{B}T^{\ast}=2\pi\hbar^{2}
n^{\frac{2}{3}}/m^{\ast}$. The mean-field self-consistent equations
are derived,
\begin{subequations}
\begin{eqnarray}
\overline n_1&=&\overline \omega t^{1/2}\Sigma_{1, \sigma=1}^{\prime}[2-D,0],\\
\label{Eq1}
1&=&\overline \omega t^{1/2}\sum_{\sigma=1,0,-1}\Sigma_{1\sigma}^{\prime}[2-D,0],\\
\label{Eq2}
\overline M_{T\neq
0}&=&t^{3/2}\sum_{\sigma}\biggl\{\Sigma_{1\sigma}^{\prime}[-D,0]
+x(g\sigma-\frac{1}{2}) \Sigma_{1\sigma}^{\prime}[2-D,0]\nonumber\\
&-&x\Sigma_{2\sigma}^{\prime}[2-D,1]\biggr\},\\
\label{Eq3} \overline \omega&=&h+4\pi\gamma \overline M,
\label{Eq4}
\end{eqnarray}
\end{subequations}
where $\gamma=q^{2}n^{1/3}/(2\pi m^{\ast}c^{2})$, and
\begin{eqnarray}\label{s1}
\Sigma_{\kappa\sigma}^{\prime}[\alpha,\delta]=\sum_{l=1}^{\infty}\frac{l^{\alpha/2}e^{-lx(\overline
\varepsilon+\delta)}}{(1-e^{-lx})^{\kappa}},
\end{eqnarray}
with $\overline \mu+g\sigma\overline \omega+2\overline
I\sigma(\overline m+\sigma \overline n_{\sigma})
=(\frac{1}{2}-\overline \varepsilon)\overline \omega$.

Similar method has been used to study the diamagnetism of the
charged spinless Bose gas \cite{Toms}. Furthermore, we have extended
it to investigate the magnetic properties of charged spin-1 Bose gas
\cite{Jian}.

\section{Results and discussions}

\begin {figure}[t]
%\vskip 0.15cm
\center\includegraphics[width=0.45\textwidth,keepaspectratio=true]{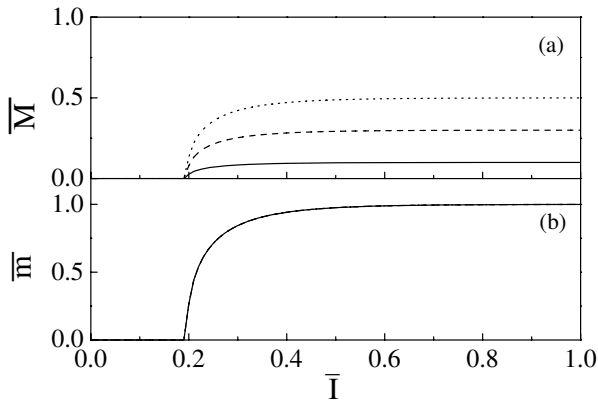}
\caption{(a) The total magnetization density $\overline M$,
(b)$\overline m= \overline n_{1}-\overline n_{-1}$ versus $\overline
I$ at reduced temperature $t=0.6$ and magnetic field $h=0.00001$.
The Lande-factor g is chosen as: $g=0.1$(solid line), 0.3(dashed
line), 0.5(dotted line).
 }
\end{figure}

In the following calculations from Fig. 1 to Fig. 6, the
characteristic parameter $\gamma$ has been set as $10^{-10}$, which
is estimated for a system with the charge and mass of $^4$He, and
the particle density being set as $(1nm)^{-3}$. Fig. 1 is plotted in
a very weak magnetic field $h=0.00001$. As shown in Fig. 1(a), the
value of total magnetization density $\overline M$ presents a
turning point from zero to nonzero. It is shown that the zero-field
spontaneous magnetization exists in this system with increasing
$\overline I$, where $\overline I$ is the reduced FM coupling of
charged spin-1 Bose gases. The curves of $\overline m$ versus
$\overline I$ in Fig. 1(b) are superposed for different
Lande-factors ($g=0.1$, 0.3 and 0.5). It suggests that $\overline m=
\overline n_{1}-\overline n_{-1}$ is independent with the
Lande-factor, so $\overline I_{c}$ at a certain temperature are
equal for any Lande-factor. Here $\overline I_{c}$ is the critical
value of reduced FM coupling of PM phase to FM phase transition.
$\overline I_{c}\approx0.19$ in this situation. When $\overline
I<\overline I_{c}$ $\overline m$ equals 0, and the value of
$\overline m$ increases with increasing $\overline I$ while
$\overline I>\overline I_{c}$ until saturate. In the region of
$\overline I>\overline I_{c}$, the magnetization density $\overline
M$ increases with Lande-factor for fixed $\overline I$, which is
attributed to the PM effect \cite{Jian}. Diamagnetism, paramagnetism
and ferromagnetism compete with each other in such system. The
diamagnetism of charged Bose gases, which is due to the internal
field induced by the spontaneous magnetization, cannot overcome
ferromagnetism in very weak magnetic field. While the competition
between paramagnetism and diamagnetism has been discussed in Ref.
13.

\begin {figure}[t]
%\vskip 0.15cm
\center\includegraphics[width=0.45\textwidth,keepaspectratio=true]{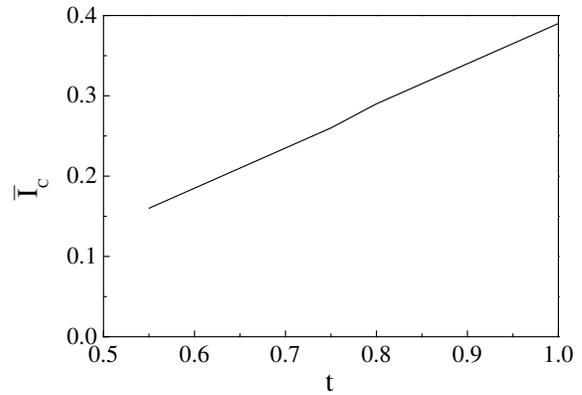}
\caption{$\overline I_{c}$ vs reduced temperature $t$ phase diagram
of charged spin-1 Bose gases at magnetic field $h=0.00001$.}
\end{figure}

Fig. 2 plots the $\overline I_{c}$ dependence of temperature at
magnetic field $h=0.00001$. The region below $\overline I_{c}$ is PM
phase, while the region above it is FM phase. As the temperature
increases, $\overline I_{c}$ increases monotonically. It is shown
that spontaneous magnetization is hard to occur at high temperature,
when the Bose statistics reduces to Boltzmann statistics.

\begin {figure}[t]
%\vskip 0.15cm
\center\includegraphics[width=0.45\textwidth,keepaspectratio=true]{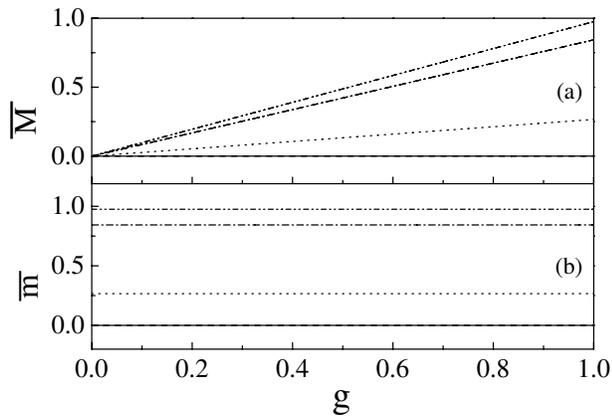}
\caption{(a) The total magnetization density $\overline M$,
(b)$\overline m= \overline n_{1}-\overline n_{-1}$ as a function of
Lande-factor $g$ of charged spin-1 Bose gases at reduced temperature
$t=0.6$ and magnetic field $h=0.00001$. The reduced FM coupling
$\overline I$ is chosen as: $\overline I=0$(solid line), 0.1(dashed
line), 0.2(dotted line), 0.3(dash dotted line), and 0.5(dash dot
dotted line).}
\end{figure}

It is supposed that $\overline m$ will reach to a nonzero
equivalence at $\overline I=0.2$ for arbitrary value of Lande-factor
for the situation of Fig. 1. To further study the influence of FM
coupling to spontaneous magnetization, Fig. 3 is plotted. It is
shown when $\overline I<\overline I_{c}(\approx0.19)$, the value of
$\overline m$ will be zero for any Lande-factor values. So the
evolution of $\overline m$ with Lande-factor $g$ are superposed and
keeps zero for $\overline I=0$ and $\overline I=0.1$. For fixed
$\overline I$ when $\overline I>\overline I_{c}(\approx0.19)$, the
magnetization density $\overline M$ increases monotonically with
$g$. While $\overline m$ maintains a constant in despite of $g$. Our
results also show that diamagnetism gives little contribution to the
magnetism in the weak magnetic field, while paramagnetism and
ferromagnetism play significant roles in the magnetization density
in the region for $\overline I>\overline I_{c}$. The interaction
between paramagnetism and ferromagnetism is intricate. The increase
of $\overline m$ due to increasing the reduced FM coupling
$\overline I$ will contribute to the paramagnetism.

\begin {figure}[t]
%\vskip 0.15cm
\center\includegraphics[width=0.45\textwidth,keepaspectratio=true]{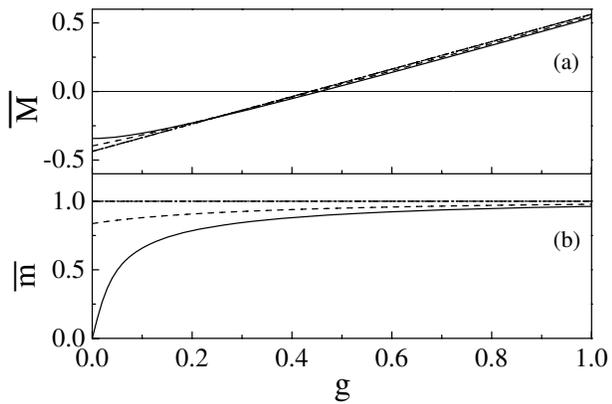}
\caption{(a) The total magnetization density $\overline M$,
(b)$\overline m= \overline n_{1}-\overline n_{-1}$ as a function of
Lande-factor $g$ of charged spin-1 Bose gases at reduced temperature
$t=0.1$ and magnetic field $h=0.1$. The reduced FM coupling
$\overline I$ is chosen as: $\overline I=0$(solid line), 0.01(dashed
line), 0.1(dotted line), 0.3(dash dotted line), and 0.5(dash dot
dotted line).}
\end{figure}

Above we have discussed the very weak magnetic field situation, now
we turn to investigate the magnetic properties of charged spin-1
Bose gases at finite magnetic field, where diamagnetism will emerges
clearly. The result of the dependence of the total magnetization
density $\overline M$ and $\overline m= \overline n_{1}-\overline
n_{-1}$ with Lande-factor $g$ at a definite magnetic field $h=0.1$
at reduced temperature $t=0.1$ is shown in Fig. 4. At low
temperature in the definite magnetic field, there is a competition
among the paramagnetism, diamagnetism and ferromagnetism. It is
shown that diamagnetism dominates in the small $g$ region, and
therefore the magnetization density exhibits negative value. When
$g>0.45$, the system presents paramagnetism which is independent of
reduced FM coupling $\overline I$. As seen from Fig. 4, the curves
of $\overline I=0.1$, $\overline I=0.3$ and $\overline I=0.5$ match
together. It means that $\overline m$ tends to saturate if
$\overline I$ is greater than a critical value. The increase of
$\overline I$ after this critical value does not contribute to the
magnetization density. Then the system exhibits similar
magnetization density at $\overline I=0.1$, $\overline I=0.3$ and
$\overline I=0.5$.

\begin {figure}[t]
%\vskip 0.15cm
\center\includegraphics[width=0.45\textwidth,keepaspectratio=true]{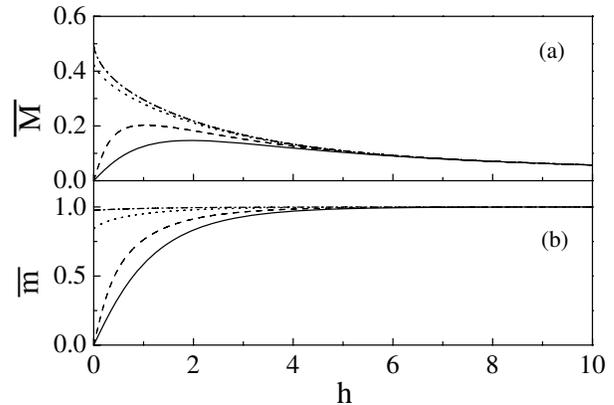}
\caption{(a) The total magnetization density $\overline M$,
(b)$\overline m= \overline n_{1}-\overline n_{-1}$ as a function of
magnetic field $h$ of charged spin-1 Bose gases at reduced
temperature $t=0.6$ with Lande-factor $g=0.5$. The reduced FM
coupling $\overline I$ is chosen as: $\overline I=0$(solid line),
0.1(dashed line), 0.3(dotted line), and 0.5(dash dotted line).}
\end{figure}

The discussions above all focused on fixed magnetic field. Next we
study the influence of magnetic field on magnetism. The evolution of
the total magnetization density $\overline M$ and $\overline m=
\overline n_{1}-\overline n_{-1}$ with magnetic field at reduced
temperature $t=0.6$ with $g=0.5$ is shown in Fig. 5. The gas always
manifests paramagnetism no matter what the values of $\overline I$
are. It indicates that in the case of $g=0.5$, diamagnetism can not
overcome paramagnetism no matter how strong the magnetic field is.
This behavior is qualitatively consistent with the result of charged
spin-1 Bose gases \cite{Jian}. In this region, the stronger
ferromagnetism induce larger $\overline m$, which will enhance
paramagnetism. With increasing the magnetic field, diamagnetism also
increases. While this will not change the paramagnetism of this
system. Whether diamagnetism can increase infinitely with magnetic
field is an important issue.

\begin {figure}[t]
%\vskip 0.15cm
\center\includegraphics[width=0.45\textwidth,keepaspectratio=true]{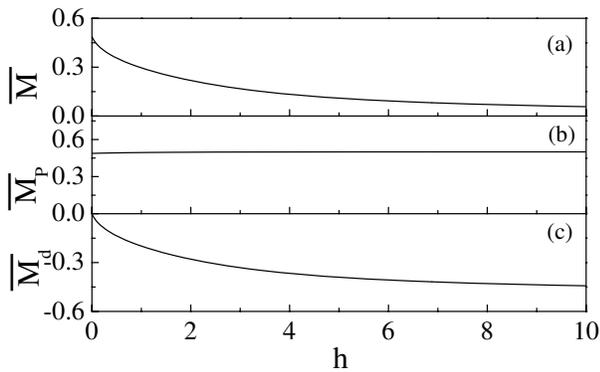}
\caption{(a) The total magnetization density $\overline M$, (b) the
paramagnetization density $\overline {M}_{p}$, and (c) the
diamagnetization density $\overline {M}_{d}$ as a function of
magnetic field $h$ of charged spin-1 Bose gases with $g=0.5$ and
$\overline I=0.5$, at reduced temperature $t=0.6$.}
\end{figure}

In order to manifest the paramagnetism and diamagnetism in detail,
in Fig. 6 we study the dependence of the total magnetization density
$\overline M$, the paramagnetization density $\overline {M}_{p}$ and
the diamagnetization density $\overline {M}_{d}$ with magnetic field
in reduced temperature $t=0.6$ with $g=0.5$ and $\overline I=0.5$.
$\overline {M}_{p}$ holds a constant since FM coupling is larger.
$\overline {M}_{d}$ tends to saturate with magnetic field. It
indicates that diamagnetism will not increase infinitely with
magnetic field. This is why in Fig. 5 the gas preserves
paramagnetism even though the magnetic field is large.

It is significant to evaluate the diamagnetic behavior at high
magnetic field limit. Without consideration of spin, the
diamagnetization density,
\begin{eqnarray}\label{lim1}
\overline M_{d}&=&t^{3/2}\sum_{l=1}^{\infty}\frac{l^{-3/2}e^{-l(\overline \omega/2-\overline \mu)/t}}{(1-e^{-l\overline \omega/t})}\nonumber\\
&\times&[1+l\overline \omega(-\frac{1}{2}-\frac{e^{-l\overline
\omega/t}}{1-e^{-l\overline \omega/t}})/t],
\end{eqnarray}
when $\overline \omega\rightarrow\infty$, $\overline M_{d}$ can be
reduced to,
\begin{eqnarray}\label{lim2}
\overline M_{d}^{\overline
\omega\rightarrow\infty}=-\frac{1}{2}\overline \omega
t^{1/2}\sum_{l=1}^{\infty}\frac{l^{-1/2}e^{l\overline
\mu/t}}{e^{l\overline \omega/(2t)}},
\end{eqnarray}
from equation (14b), we can obtain,
\begin{eqnarray}\label{lim3}
1=\overline \omega
t^{1/2}\sum_{l=1}^{\infty}\frac{l^{-1/2}e^{l\overline
\mu/t}}{e^{l\overline \omega/(2t)}},
\end{eqnarray}
Substituting equation (\ref{lim3}) into (\ref{lim2}), $\overline
M_{d}^{\overline \omega\rightarrow\infty}=-1/2$ can be obtained.
This analytical result illustrate the diamagnetization density
$\overline {M}_{d}$ tends to a finite value at high magnetic field.

\begin {figure}[t]
%\vskip 0.15cm
\center\includegraphics[width=0.45\textwidth,keepaspectratio=true]{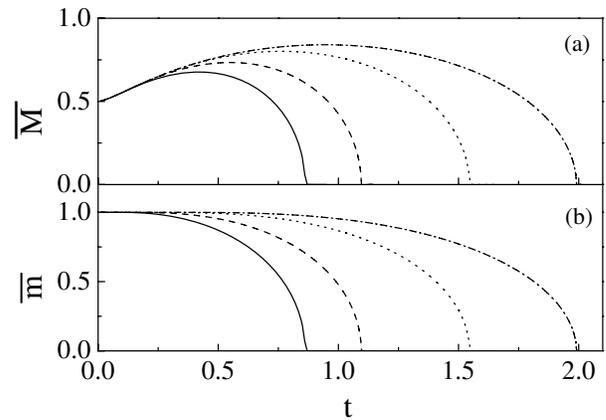}
\caption{(a) The total magnetization density $\overline M$,
(b)$\overline m= \overline n_{1}-\overline n_{-1}$ versus reduced
temperature $t$ of charged spin-1 Bose gases with $\gamma=0.1$ and
$g=1$, at magnetic field $h=0.00001$. The reduced FM coupling
$\overline I$ is chosen as: $\overline I=0$(solid line), 0.1(dashed
line), 0.3(dotted line), 0.5(dash dotted line).}
\end{figure}

In order to investigate the magnetic properties of the charged
spin-1 Bose gas in low temperature, we suppose $\gamma=0.1$. The
evolution of the total magnetization density $\overline M$ and
$\overline m= \overline n_{1}-\overline n_{-1}$ with reduced
temperature at $h=0.00001$ and $g=1$ is shown in Fig. 7. It is shown
that $\overline M$ increases with increasing temperature, and
reaches a maximum, then decreases at high temperature region. The
upward trend at low temperature reflects the diamagnetism, comparing
with our results in Ref. 21, which shows a flat trend at the same
temperature region. A sharp decline can be seen when $\overline M$
close to zero. This suggests that there is a pseudo-condensate
temperature in the transition from ferromagnetism to paramagnetism.
Although condensation has not been considered, the magnetic field is
faint in such a case. It is reasonable that the pseudo-critical
temperature increases with increasing reduced FM coupling $\overline
I$. Therefore, the temperature region of ferromagnetism enlarges
from $\overline I=0$ to $\overline I=0.5$ in turn.

\section{Summary}

In summary, we study the interplay among paramagnetism, diamagnetism
and ferromagnetism of charged spin-1 Bose gas with FM coupling
within the mean-field theory. In very weak magnetic field, it is
shown that the ferromagnetism is stronger than the diamagnetism,
where the diamagnetism is related with spontaneous magnetization.
The critical value of reduced FM coupling $\overline I_{c}$ of PM
phase to FM phase transition increases with increasing temperature.
The Lande-factor $g$ is supposed as a variable to evaluate the
strength of the PM effect. The gas exhibits a shift from
diamagnetism to paramagnetism as $g$ increases at a finite magnetic
field. Ferromagnetism plays an important role in the magnetization
density in the weak magnetic field. Diamagnetism can not increase
infinitely with magnetic field at high magnetic field. Condensation
is predicted to occur through studying the low-temperature magnetic
properties in a weak magnetic field.

\section*{Acknowledgments}
JQ would like to thank Professor Huaiming Guo for the helpful
discussions. This work was supported by the National Natural Science
Foundation of China (Grant No. 11004006), and the Fundamental
Research Funds for the Central Universities of China.

\end{document}